\documentclass[conference]{IEEEtran}
\ifCLASSINFOpdf
\else
\fi
\usepackage{soul}
\usepackage{rotating}
\usepackage{longtable}
\usepackage{booktabs}

\usepackage[table]{xcolor}
\usepackage{graphicx}
\usepackage{caption}
\usepackage{tabularx}

\usepackage{wrapfig}
\usepackage{amsmath}
\usepackage{graphicx}
\usepackage{caption}
\usepackage{footnote}

\usepackage[numbers]{natbib}

\begin{document}
%
\title{Cloud Service Matchmaking Approaches: A Systematic Literature Survey}

\author{\IEEEauthorblockN{Beg{\"u}m {\.I}lke Zilci, Mathias Slawik, and Axel K{\"u}pper}
\IEEEauthorblockA{Service-centric Networking\\Technische Universit{\"a}t Berlin, Germany \\
\{\textit{ilke.zilci}\textbar \textit{mathias.slawik}\textbar \textit{axel.kuepper}\}@\textit{tu-berlin.de}}
}


%


\maketitle

\begin{abstract}

Service matching concerns finding suitable services according to the service requester's requirements, which is a complex task due to the increasing number and diversity of cloud services available. Service matching is discussed in web services composition and user oriented service marketplaces contexts. The suggested approaches have different problem definitions and have to be examined closer in order to identify comparable results and to find out which approaches have built on the former ones. One of the most important use cases is service requesters with limited technical knowledge who need to compare services based on their QoS requirements in cloud service marketplaces. Our survey examines the service matching approaches in order to find out the relation between their context and their objectives. Moreover, it evaluates their applicability for the cloud service marketplaces context.
\end{abstract}


%
\IEEEpeerreviewmaketitle










\section{Introduction}
The increasing number and diversity of services available in the cloud computing environment causes the service selection to become more complex. Cloud service marketplaces in the industry provide merely keyword search and categories to select applications from such as marketing and customer management \cite{googlemarketplace}, \cite{appexchange}. Moreover, in multi-cloud application management platforms such as Right Scale or Bitnami service requesters have the option to deploy parts of their applications on different clouds. Service matching research aims at finding suitable services according to the service requester's requirements. These requirements cover a wide range of properties such as functional, quality of service (QoS) and compliance.   

The service matching problem is discussed in the literature in different contexts: Internet of Services, SOA, Semantic Web Services and Cloud Services. For service descriptions, several languages are suggested, each focusing on another group of service properties depending on the context such as WSDL\cite{christensen2001web}, Linked-USDL\footnote{Linked Unified Service Description Language} \cite{linkedUSDL}, SMI\footnote{Service Measurement Index} \cite{smi2014v2} and OWL-S\footnote{Semantic Markup for Web Services} \cite{martin2004owl}. Based on the context, the \textit{scope of matching} covers QoS matching, input-output matching, precondition and postcondition matching. 

Due to different opinions on the content of the service descriptions and different contexts, a variety of service matching approaches have been suggested. However, these approaches have different problem definitions and have to be examined in detail in order to identify the subproblems solved in each.



Research questions that this survey attempts to answer are:

\textbf{Q1} How do they select QoS parameters to work with? Based on interviews, non-functional properties research or datasets? 

\textbf{Q2} Where do the service descriptions stem from: cloud computing or web services context? 

\textbf{Q3} What context do they target in their problem definitions? Web services, semantic web services or cloud computing in particular?

\textbf{Q4} Are there any approaches which meet the essential requirements of cloud service marketplaces?

\textbf{Q5} Do the existing approaches support all data types to be matched?

Our survey builds on the related work presented in \cite{platenius2013survey} and \cite{jula2014cloud}. These surveys provide a good overview of service matching approaches, however they have different focuses, on-the-fly computing and cloud service composition, respectively. 

To answer our research questions \textbf{Q1}, \textbf{Q2}, and \textbf{Q3} we examine the service description languages used, the source of service descriptions and the research or project context where the matching approach was developed. For \textbf{Q4} and \textbf{Q5}, we assess to what extent the service matching approaches accomplish their objectives with an in-depth examination of the most relevant approaches for the context where the user takes the final decision of service selection. Moreover, we provide clear statements on the relation between the objectives and the service matching context.


This paper is organized as follows: The next section presents the survey procedure. Section 3 examines the existing approaches to answer the research questions. Section 4 presents the discussion.



\section{Survey Procedure}

Within ACM, IEEE, and Springer digital libraries, we have searched with the keywords "service composition", "service matching", "service matchmaking", "service selection", and "QoS". Following the references, we added some more to the list. We omitted the approaches which do not fully explain how to perform QoS matching or provide only initial steps. Following this, we evaluated the approaches with two essential requirements in the cloud service marketplaces context: service requester perspective and handling incomplete knowledge. Section \ref{sec:existingApp} examines the related work to answer the research questions \textbf{Q1-Q5}.




\section{Existing Service Matching Approaches}\label{sec:existingApp} 
As service matchmaking approaches evolve, several methods and toolsets are suggested and evaluated in the literature. This section examines existing approaches closely.

Table \ref{tab:toolsettable} shows the examination of matching approaches to answer the research questions provided above. Each column is described as follows: 

\textbf{Service Descriptions Source } lists the service description language used for describing the services or the source for the QoS parameters in the service descriptions. This information is utilized to answer \textbf{Q1} and \textbf{Q2}.


\textbf{Target Context} explains for which context the matching approach was designed, since the research objectives might vary accordingly. Moreover, the target features of the matcher depend on the analysis of the QoS properties to be matched. This field targets answering \textbf{Q3}.

\textbf{Calculation concepts} refer to the mathematical model(s) used to structure the problem.\footnote{FL=Fuzzy Logic MIP= Mixed-Integer Programming CP=Constraint 
Programming MMKP,MCOP=Multidimensional Multiple-choice 0-1 Knapsack Problem, Multi-Constrained Optimal Path FL=Fuzzy Logic AHP=Analytic 
Hierarchy Process NFP=Non-Functional Property IFS=Intuitionistic Fuzzy Sets IOPE= Input, Output, Precondition, Effect FL=Fuzzy Logic FST=Fuzzy Set Theory SR=Subsumption Reasoning PCM=Policy Centered Meta-Model SWRL=Semantic Web Rule Language IFS=Intuitionistic Fuzzy Sets SWS= Semantic Web Services}

Three main sources for the service descriptions are: i) Numerical QoS parameters, non-functional properties research by \citeauthor{al2007qos} \cite{al2007qos} and by \citeauthor{menasce} \cite{menasce} in web services context, examples include execution time, price, availability, ii) WSDL input and output parameters, iii) Semantic web service challenge and OWL-S. The specific requirements of service matchmaking in cloud computing is to be further investigated. Apart from the contributions by the developers of USDL and Linked USDL \cite{linkedUSDL}, and SMI \cite{garg2013framework} themselves, there are no matching approaches which are applied to these cloud context non-functional properties.

\begin{table*}[htp]
\rowcolors[]{3}{white}{gray!30}
\rotatebox{360}{ 
\begin{minipage}{\textheight} 
\begin{tabular}{p{30mm}|p{40mm}p{40mm}p{40mm}}    \toprule 
\textbf{Research Work} & \textbf{Properties} &&  \\\midrule 
\textbf{Matcher}   		 & \textbf{Service Descriptions Source} 		& \textbf{Target Context} & \textbf{Calculation Concepts}  \\ 
\hline
\citeauthor{yu2007efficient} \cite{yu2007efficient}  &	 numeric QoS parameters, cites non-functional properties research in web services context with focus on performance and cost optimization by Menasce &	web services composition in SOA &	 MMKP, MCOP 	 \\
\citeauthor{hu2008web} \cite{hu2008web}  & 	numeric QoS parameters by Menasce & web services composition in SOA &	 vector normalization  \\
\citeauthor{kritikos2009mixed} \cite{kritikos2009mixed}  &	numeric QoS parameters along with their measurement units and methods in OWL-Q\cite{kritikos2006semantic}, OWL-S & 	WS & 	 MIP \\ 
\citeauthor{kritikos2007semantic} \cite{kritikos2007semantic} & 	 OWL-Q\cite{kritikos2006semantic}, OWL-S &	 WS &	 CP \\
\citeauthor{jie2011dynamic} \cite{jie2011dynamic} &	 SWRL rules, OWL-S, WSDL &	 semantic WS, sws-challenge.org &	 - \\ 
\citeauthor{palmonari2009effective} \cite{palmonari2009effective} &	WSMO, WSML &	 semantic WS &	 -  \\ 
\citeauthor{d2008semantic} \cite{d2008semantic} &	 OWL-S &	 semantic WS in IoS, e-shopping domain &	 SR 	 \\ 
\citeauthor{sarang2012clustering} \cite{sarang2012clustering} &	WSDL input output parameters &	 WS from IBM UDDI registries	& Peano space filling curve	\\ 
\citeauthor{eleyan2011service} \cite{eleyan2011service} & as described in \cite{eleyan2010extending}&	 WS &	 AHP and \newline Euclidean distance \\
\citeauthor{mobedpour2013user} \cite{mobedpour2013user} & numeric QoS with QWS dataset \cite{al2007qos},\cite{al2007discovering}  &  	 WS &	 FL  \\ 
\citeauthor{wang2009qos} \cite{wang2009qos}	& user perception &	service marketplace &	IFS,FL 	 \\ 
\citeauthor{bacciu2010adaptive} \cite{bacciu2010adaptive}&numeric QoS parameters & service composition	& FST,FL  \\ 
\citeauthor{liu2009weighted} \cite{liu2009weighted}, \cite{liu2012fuzzy} &	 WSDL + OWL-S,WSMO &	 semantic WS &	 SR, FL 	 \\ 
\bottomrule 
\end{tabular}
\end{minipage} } 
\caption{Examination for Service Descriptions Used} \label{tab:toolsettable} \end{table*}%
Table \ref{tab:basicReqTable} aims at answering \textbf{Q4}.
It evaluates the approaches with two essential requirements in the cloud service marketplaces context. 
\textbf{Non-technical service requester perspective} assesses if the matcher answers one or all the needs of non-technical requesters with QoS requirements by allowing them to: i) make the final decision, ii) set weights on each constraint, iii) specify constraints for each property which will be compared to the service specifications rather than arranging the repository in itself, iv) specify fuzzy requests.  
\textbf{Incomplete knowledge}
refers to how the matchmaking method deals with missing constraints in the service query and missing properties in the service description. 

In the list provided in the survey by \citeauthor{platenius2013survey} \cite{platenius2013survey}, the QoS matching approaches which explicitly deal with incomplete knowledge are: \cite{palmonari2009effective}, \cite{wang2009qos}, and \cite{bacciu2010adaptive}. Some approaches provide gradual results which implicitly deal with it: \cite{d2008semantic}, \cite{liu2009weighted}, and \cite{liu2012fuzzy}.

\begin{table*}[htpb]  
\rowcolors[]{3}{gray!30}{white}
\begin{minipage}{\textheight} 
\begin{tabular} {p{30mm}|p{60mm}p{50mm}p{10mm}}   
  \toprule \textbf{Research Work} & \textbf{Properties} &&  \\\midrule 
   \textbf{Matchmaker}    &  \textbf{Service Requester Perspective}  & \textbf{Incomplete Knowledge \newline(requester-provider)} &\textbf{Adequate}\\ 
\hline
\citeauthor{yu2007efficient} \cite{yu2007efficient}   & no, intended for dynamic service composition & works with precise data & no   \\
\citeauthor{hu2008web} \cite{hu2008web}  & yes, weights on constraints & no & no \\
\citeauthor{kritikos2009mixed} \cite{kritikos2009mixed} & yes, constraints for each property & yes,classified results & yes \\
\citeauthor{kritikos2007semantic} \cite{kritikos2007semantic} & yes, classifies according to requester's needs& implicitly yes, two of the algorithms return classified results& yes  \\
\citeauthor{jie2011dynamic} \cite{jie2011dynamic}  & performs comparison of requesters constraints to services&not explicitly documented &no\\
\citeauthor{palmonari2009effective} \cite{palmonari2009effective} & yes, but how QoS request will be converted WS-Policy or WSLA and to PCM is not documented & both on the requester and the provider side & no   \\
\citeauthor{d2008semantic} \cite{d2008semantic} & merely ordering service descriptions, no comparison to service requester's constraints
& implicitly yes, returns gradual results 
using semantic distances & no  \\ 
\citeauthor{sarang2012clustering} \cite{sarang2012clustering}  &no, clusters services with similar ones &not explicitly documented & no  \\ 
\citeauthor{eleyan2011service} \cite{eleyan2011service}  & yes, uses AHP \footnote{Analytic Hierarchy Process} for converting the service requester’s preferences to weights & implicitly yes,ranks services in preference order& yes   \\ 
\citeauthor{mobedpour2013user} \cite{mobedpour2013user} & yes, user-centered & implicitly yes, returns super-exact and partial matches & yes \\ 
\citeauthor{wang2009qos} \cite{wang2009qos} & yes, addresses imprecise preferences of service requesters & on the provider side  & yes   \\
\citeauthor{bacciu2010adaptive} \cite{bacciu2010adaptive} & yes, fuzzy sets to allow imprecise service requests & on the provider side & yes  \\
\citeauthor{liu2009weighted} \cite{liu2009weighted}, \cite{liu2012fuzzy} &yes, constraints for each property &gradual results&yes\\

 \bottomrule \hline \end{tabular}\end{minipage}  \caption{Evaluation for Essential Requirements of Cloud Service Marketplaces} \label{tab:basicReqTable}
\end{table*}   

Table \ref{tab:summaryTable} aims at answering  \textbf{Q5}. Some approaches only explain how the format of the constraint should be, however they do not state how the service descriptions should be. Mostly, the QoS property is described as a tuple that can be added to existing service description languages. 

\begin{table*}[htpb] 
\rowcolors[]{3}{gray!30}{white}
 \begin{minipage}{\textheight} 
 \begin{tabular} 
{p{20mm}|p{35mm}p{25mm}p{30mm}p{35mm}}    \toprule 
\emph{Research Work} & \emph{Properties} &&&  \\\midrule 
Matchmaker    &  QoS Constraint\newline Description & Service\newline Description &Data Types of Constraints & Prototype \\ 
\hline
\cite{eleyan2011service} & as described in \cite{eleyan2010extending} & WSDL extension & numeric & In C\# \\  
\cite{mobedpour2013user}  & QoS query language which separates relaxation and preference orders\cite 
{mopedpour2010lang} & QWS dataset\cite{al2007qos},\cite{al2007discovering} & fuzzy \footnote{only for numeric properties such as reliability, response time}, range, discrete numeric data & 
non-functional matching layer: C\#, ASP.NET, lp\textunderscore solve \footnote{functional matching layer: Lucene and Java}\\ 
\cite{bacciu2010adaptive} & Fuzzy 
Sets & Trapezoidal fuzzy numbers & Only Numeric Examples
& Extension in MATLAB, Dino Broker in Java\\ 
\cite{wang2009qos} & linguistic terms & no formal description & decision maker's perception 
converted to intuitionistic fuzzy number \footnote{this one addresses non-measurable QoS properties which the author calls non-functional properties 
such as reliability, \newline integrity,availability and security} & -\\
 \cite{kritikos2009mixed},\cite{kritikos2007semantic}
& as described in \cite{kritikos2006semantic} & as described in \cite{kritikos2006semantic} & numeric, range & using MATLAB and Java \\ \bottomrule 
 \hline \end{tabular} \end{minipage} 
 \caption{Summary of the Most Relevant Related Work} \label{tab:summaryTable} 
 \end{table*}

The work by \citeauthor{eleyan2010extending} \cite{eleyan2010extending} states that non-functional properties should be added to the service matching process, however they do not
specify what they improve from other matching approaches that handle non-functional properties. Moreover, they assume the service descriptions to be equal to the query in terms of properties they include. 
 
The aim of \citeauthor{wang2009qos} \cite{wang2009qos} is to develop a matchmaking approach for QoS properties which cannot be measured and depend largely on the 
perception of service providers and consumers. It models the problem as a fuzzy multi-criteria decision making problem. The proposed method takes 
linguistic terms which express the decision maker's opinion on the QoS property as input; e.g., very good, fair, poor. These linguistic terms are 
converted to intuitionistic fuzzy numbers, and scores for each service based on decision maker's weighting for each property are calculated. Finally, 
the results of the decision makers are aggregated, and the services are ordered from the best to the worst.  

\citeauthor{mobedpour2013user} \cite{mobedpour2013user} state that in existing matching approaches such as \cite{kritikos2009mixed} and \cite{ruiz2005improving} that the 
non-technical or inexperienced service requesters are expected to learn their service query languages or to gain background knowledge on defining fuzzy membership 
functions. To solve this problem, the authors propose a process where the service requester is supported and guided by the system to formulate QoS queries along 
with an improved selection model. In contrast to the approach by \citeauthor{wang2009qos} \cite{wang2009qos}, the authors utilize interval clustering to classify 
services under the linguistic terms, which also decreases user effort. These ideas are implemented in a prototype which uses the algorithm presented 
in \cite {kritikos2009mixed} as a baseline. The experiments show that more accurate results are achieved while the response time remains as before. 
However, no user reviews are conducted to evaluate the main purpose of this approach. Service descriptions do not contain any enumeration parameters 
and parameters related to business. This results in that the system does not support a use case where there are different price models for several 
editions of a service.  

\citeauthor{mobedpour2013user} \cite{mobedpour2013user} criticize fuzzy models approaches since they 
require high user evaluation effort (when they have to define fuzzy numbers or membership functions) and that they might be wrong since they are based 
on personal opinions of the decision makers. Moreover, they do not consider crisp data. However, the approach represented in \cite{wang2009qos} takes 
linguistic terms as input which does not require substantial user effort. On the other hand, the QoS description is based on users' personal 
impressions. The last step of aggregating the scores to rank the services improves this aspect to some extent.   

\citeauthor{bacciu2010adaptive} \cite{bacciu2010adaptive} address 
the service composition problem in two steps: integration of heterogeneous services and dynamic service selection---since the service properties change 
due to resource availability or network connectivity variations. The authors present three artifacts: a method for fuzzy specification of QoS 
parameters both in service requests and descriptions, a matchmaking procedure for those, and a method which targets the dynamic update of service 
descriptions. For doing so, they extend an existing service broker which had crisp matching functionality with the imprecise QoS parameter support. 
\citeauthor{bacciu2010adaptive} \cite{bacciu2010adaptive} report that the extension in MATLAB code is planned to be integrated to Dino Broker \cite{mukhija2007qos}, however, there are no publications 
about the integration. The available version of Dino Broker is in Java and the published code does not contain the fuzzy extension. \citeauthor{bacciu2010adaptive} \cite{bacciu2010adaptive} present an example of variational scope as a source of fuzziness. This approach expresses numeric QoS descriptions and 
requirements as trapezoidal fuzzy numbers. The authors assume that when a QoS requirement is a range such as "availability between 95\% and 99\%", 
the acceptable service specifications start at 94\% with a lower membership value, increasing to the best membership value between 96\% and 98\%, and 
the membership value starts decreasing at 98\% and ends at 100\%. The idea that neighbor values of the specified range are also likely acceptable by the service requester is correct on the lower side, 
however on the upper side it does not make sense for properties with positive tendency because it considers 99\% less good than 98\%, although availability is a high-value preferred property. The 
authors give examples for the application with numeric values, however not with enumeration type QoS descriptions such as location or compatible 
browsers, although it is mentioned that the operators of set theory can be generalized to fuzzy set theory.


\section{Discussion}

At first glance, there are cloud service matchmakers which try to solve the same service matchmaking problem. However, the problem definition 
varies according to the use cases and project contexts. For this reason, they either optimize on automation and leave the usability issues aside or vice versa. The different problem definitions result in their solutions not being directly comparable.


From our analysis in Table \ref{tab:toolsettable}, we conclude the following. In most cases, the project context provides the service description language used. If the service description language is an ontology, the service matchmaker is based on ontology 
reasoners. In other cases, the service matchmakers use different mathematical methods. However, the service matchmakers vary also because of other 
factors: the target service requesters and their definition for the process of service matchmaking.

One group is for automatic service composition, where the requester is another service and the service matchmaker has to find the best service specification for the request. Another group is user-oriented where the requester is a person who makes the final service selection from a list of alternative services. For the former group, the process of service matchmaking ends with the presentation of the optimal solution to the problem. For the latter group, further requirements are considered: (i) a categorization of results as very good, good and acceptable, (ii) providing an initial list and then applying relaxation, (iii) step-by-step guidance for the service requester to add constraints to the request. Another problem definition aims at answering "a developer's query" which focuses on matching the input output parameters of functions. An additional consideration in the related work is static and dynamic service descriptions, some approaches foresee an integrated monitoring module and updating the service descriptions accordingly (\textbf{Q3}). 

Table \ref{tab:basicReqTable} shows that there are some approaches which meet the essential requirements of cloud service marketplaces on a high level (\textbf{Q4}).



The service description languages mostly provide very generic property definitions, therefore the constructs give too much freedom to the person 
who describes the service and no standardized measures for most of the properties. While service description languages such as Linked USDL \cite{linkedUSDL} and 
Service Measurement Index \cite{smi2014v2} define properties of services in very detailed yet generic ways, most existing service matchers use only numerical 
values as examples for their proof-of-concept implementations. 



In the existing approaches, the service descriptions stem mostly from the web services context. Some QoS properties which are specific to cloud services are not considered, for example scalability, elasticity and different price models. Moreover, some matching approaches do not provide concrete examples for the service properties their service matcher targets. An example for this is the n-ary constraints in the mixed integer programming approach \cite{kritikos2009mixed}. Another example is the constraint-specification combinations with intervals and values. While they examine interval to interval exact matching in great detail (this combination comes quite rarely in our SDL and SLA parameters in general), they do not solve the feature list matching problem (\textbf{Q1,Q2}). 

In our work \cite{zilci2015cloud}, we provide an analysis of QoS properties in service descriptions. Based on this analysis, we define the subproblems of service matching as discrete numeric matching, enumeration matching, feature list matching, interval matching, and fuzziness. We examine the matching approaches presented in  \cite{mobedpour2013user}, \cite{bacciu2010adaptive} and \cite{mukhija2007qos}, the three complementary papers \cite{kritikos2009mixed},\cite{kritikos2007semantic}, and  \cite{kritikos2008evaluation}, and the approach presented in \cite{wang2009qos}. Most of the service matchers address discrete numeric matching and interval matching. Some matchers consider also enumerations: the query and the specification can take only one of the values from a predefined list. In feature list matching, both the query and the specification can take multiple values from a predefined list and a greater overlap of these two lists is targeted. However, none of the service matchers cover this subproblem (\textbf{Q5}).
 
\section{Conclusion and Future Work}
The differences in the definition of the service matching problem results in the authors defining the functional requirements for the service matcher. The tools are selected according to the requirements. The different problem definitions of the related work result in conceptual artifacts which should be carefully examined for reusing. A catalog of components which address the subproblems of service matching can be useful to identify building blocks available to be combined.


\section*{Acknowledgment}
This work is supported by the Horizon 2020 EU funded Integrated project
CYCLONE\footnote{cyclone-project.eu}, grant number 644925.

\bibliography{main}{}
\bibliographystyle{myIEEEtranN}

\end{document}